\documentstyle[eqsecnum,prd,aps,epsf,12pt]{revtex}

\begin{document}
\tighten
\title{ \bf {\large Cosmic String Helicity: {\LARGE Constraints on Loop Configurations, and the \\ Quantization of Baryon Number}}}
\author{\large\bf Tsippy R. Mendelson\thanks{E-mail:~tsippora@shum.huji.ac.il}\thanks{E-mail:~perachsheli@hotmail.com}}
\address{Racach Institute of Physics, Hebrew University, Givat Ram, Jerusalem~91904, Israel.}

\maketitle

\begin{abstract}
We apply the concept of helicity from classical hydrodynamics to elucidate
two problematical issues in cosmic string physics. Helicity, the space
integral of the scalar product of a velocity-like field with its vorticity
field (curl), can be defined for a complex scalar field in analogy with 
fluids. We dwell on the topological interpretation of helicity as related
to the linking of field lines of the vorticity field. Earlier works failed
to fully implement this interpretation for cosmic strings by missing a term 
connected with the linking of these lines inside the strings. As a result
paradoxical conclusions were drawn: global cosmic string loops may not take 
on certain simple shapes, and baryon number is not quantized in integers 
in the presence of local cosmic strings in gauge theory. We show that both
paradoxes are removed when internal contributions to helicity are properly
taken into account. In particular, quantization of baryon number can be
understood within a special case of the Glashow-Weinberg-Salam model if cosmic 
strings are the unique mechanism for baryosynthesis. In addition, we find
a new constraint on the permitted linkages of cosmic strings in a 
string tangle.    
\end{abstract}

\section{ \bf  Introduction}
Cosmic strings are a special case of  
the vorticity phenomenon, which appears in many forms - in the flow of 
rivers and streams, in smoke rings, in hurricanes and tornadoes, etc. A special
case of vorticity are the \textit{quantum vortices}~\cite{onsager,feynman_55} which include vortices 
in superfluids, flux tubes in superconductors and cosmic strings. Global cosmic strings  are analogues to Vortices in superfluids while local cosmic strings   are analogues to flux tubes in superconductors~\cite{vilenkin_94}. 
All these vortices are described by a complex wave function or scalar field
with a multivalued phase function  $\theta$, such that
\begin{equation}
\label{q_phase}
\oint \nabla \theta \cdot d{\bf l} = 2 \pi n \quad \quad n = \pm1,\pm2,...
\end{equation} 
over a closed contour surrounding the string. As a result different physical quantities get quantized: the 
circulation of a superfluid vortex, the magnetic flux in a superconductor, the 
magnetic flux in a local cosmic string.

Vorticity was first investigated by Helmholz and Kelvin who set 
down the foundations for the study of vorticity in hydrodynamics. 
In this work we
try to implement concepts born in the field of classical hydrodynamics to
cosmic strings. As we shall see the interaction between these two fields
can be very rewarding. Turbulent fluid is regarded as full of vortex filaments~\cite{moffatt_81}. Quantum turbulence in superfluids has likewise been defined
as a superfluid state featuring a tangle of quantum vortex filaments~\cite{donn_86}. This paper is concerned with helicity of a tangle of cosmic strings. 

The concept of \textit{fluid helicity} was introduced by Moffatt~\cite{moffatt_69} as a useful measure of the degree of linkage of ordinary fluid vortice loops. It is defined as the volume integral of the scalar product of a velocity field $\bf v$ and its curl, the vorticity $\bf \xi$. Following this, Moffatt~\cite{moffatt_81,moffatt_69,moffatt_92} and others~\cite{berger_84,arnold_74,kuznetsov_80} considered  helicity integrals in general, i.e. the space 
integral over the scalar product of a vector field and its curl
\begin{equation}
H = \int {\bf V}\cdot \nabla \times {\bf V} \, d^3r,
\end{equation}  
and found them to 
have a topological interpretation as the linking of the field lines of the divergence free field $\nabla \times {\bf V}$. Thus the conservation of helicity can
be explained for systems in which the field lines may not cross each other,
 such
as vortex lines in an inviscid fluid and the magnetic field lines in a 
perfectly conducting fluid~\cite{moffatt_69}. For these flows the field lines are said to be
frozen into the flow, and the helicity is a topological invariant  
mathematically related to the Hopf invariant~\cite{hopf_31} classifying the nontrivial 
homotopy classes of maps from $S^3$ to $S^2$.\footnote{The existence of a 
relation between 
helicity integrals and the topological Hopf invariant has been mentioned 
by~\cite{moffatt_81,berger_84,kuznetsov_80,whitehead_47}. However, the exact
relation had been derived by Arnold~\cite{arnold_74}. The Hopf invariant 
has an elementary geometric interpretation as the linking number of the preimages of the points of the target space $S^2$, which are isomorphic to circles in
$S^3$. The Hopf invariant is equivalent to the helicity integral of a vector
field who's field lines lie tangent to the preimage circles. In this sense
the Hopf invariant is a helicity integral term, but the inverse is usually not
the case.}

Superfluids and cosmic strings are both described
by field theories of a complex field
$\psi = \rho e^{\imath \theta}$ which obeys a nonlinear
field equation~\cite{davis_89}, e.g. the nonlinear Ginzburg-Pitaevskii equation~\cite{Ginzburg_58} or the Higgs equation. Cosmic strings and superfluid vortices both occur 
in the presence of spontaneous symmetry breaking~\cite{kibble_76}. 
A vortex filament or cosmic
string is a configuration of $\psi$ which approaches asymptotically the
broken symmetry vacuum, and is characterized by a phase $\theta$ which changes
by $2\pi N$ ($N \in {\bf Z}$) when one goes once around the filament or string axis.
$N$, is termed the \textit{winding number of the
string}. 
Single valuedness of the field $\psi$ on the filament axis requires that $\rho = 0$ along it. The curves along which $\rho = 0$ define the
position of the strings. A vortex filament or cosmic string must either be infinite in both directions, close on itself (cosmic string loop, vortex ring), or terminate on a 
boundary
of the system. Otherwise the behavior of 
the phase just beyond the filament's free ends would be ambiguous. We assume throughout that all
strings are closed un-knotted loops confined to a finite region. 

The analogy between vortices in
superfluids and global cosmic strings has inspired the construction of a \textit{helicity} for global cosmic strings.
 Bekenstein~\cite{bekenstein_92} defined a helicity for global strings, and Vachaspati and Field~\cite{vachaspati_94}, Sato and Yahikozawa~\cite{sato_95a} and Sato~\cite{sato_95b} did this for local strings.
These works tried to relate helicity for cosmic strings
with the topological and geometrical structure of these strings (such as linking, knotting, writhing and twisting). In all these works
bizarre physical conclusions were drawn. Bekenstein concluded that an isolated unknotted loop of global string is restricted to a plane~\cite{bekenstein_92} i.e., only planar configurations may exist. This is strange since dynamically there seems to be no constraint on a loop accumulating a distortion continuously in an arbitrary direction. Furthermore simulations of the formation and evolution of cosmic strings show the formation of loops which are not planar~\cite{vachaspati_84,shellard_90}.
Bekenstein also claimed that the contortion of a single knot is quantized in integers. This is also strange because it says a single knot may not wiggle freely about, but is frozen in a configuration of integer contortion.

 For local cosmic strings there is a relation between the electroweak magnetic helicity of a tangle of strings and the violation of baryon number conservation via what is known as the chiral anomaly~\cite{vachaspati_94,sato_95b}.
Vachaspati and Field, Sato and Yahikozawa and Sato found the helicity 
to change continuously with the shape of the strings, which implies that baryon number changes continuously too.
This clashes with the expectation that baryon number is quantized in integers.

 We show here that the topological interpretation of helicity 
as the
linking of field lines themselves is not fully incorporated in these works, and hence an 
important term was left out, which accounts for all the mentioned unacceptable
conclusions. Helicity arises from internal structure  
within a string, determined by the twist and writhe of the string, in addition to the external
relations between strings, i.e. linking and knotting. 
The missing term arises from the internal structure of field lines
within the strings which we name \textit{internal helicity}.
 It was previously believed~\cite{vilenkin_94} that since the strings length is much greater than its width, the internal structure of the string becomes unimportant and physical quantities of interest can be averaged over the string cross-section.
However, as we will see this is not always the case.

In order to clarify the contribution of internal helicity, the
relation of helicity to the topological structure of field 
lines will be further developed here. Field lines of vector fields which 
are divergence-free do
not have endpoints. This property allows us to examine such field structures
in terms of the topology of closed curves. The link between
helicity and topological invariants of curves was first conjectured by 
Moffatt in 1981~\cite{moffatt_81} and then developed by Berger and Field in 1984~\cite{berger_84}. In 1992
Moffatt and Ricca~\cite{moffatt_92} managed to derive the extra term, arising from the internal structure of magnetic flux tubes, directly from the
field equations of motion for a magnetohydrodynamic fluid. This is not possible when calculating the helicity of cosmic strings since their equations are second order non-linear equations, and therefore the calculation is too complicated. However, using formulas describing curves, developed by mathematical biologists investigating the structure of DNA~\cite{white_69,fuller_71},
 the internal helicity may be calculated.    
By adding it to the external helicity we correct the mentioned results.

 In section~\ref{sec_2} we review the concept of \textit{helicity} of a solenoidal (divergence free) vector field, and dwell on the relation of the helicity to the linking of the field lines of the solenoidal field. 
In section~\ref{sec_3} we review the construction and calculation of the helicity for a tangle of global cosmic string loops, as a function of different geometrical and topological properties
 of the strings~\cite{bekenstein_92}. Next we correct the previously mentioned
conclusions of Bekenstein by means of the internal helicity.
In section~\ref{sec_4} we deduce new constraints on the permitted linkages of string
loops. We find that the linking possibilities of string loops depends on their winding numbers.
 In section~\ref{sec_5} we review how the electroweak helicity of local cosmic strings in the Weinberg-Salam model is related to the baryon number, and clarify how baryon number conservation is violated. The baryon number is found to be quantized, in contrast to previous results. Again, we show how internal helicity 
accounts for the new results.

\section{\bf  Helicity as Topological Invariant}
\label{sec_2}
The term \textit{helicity} is used in particle physics for the scalar
product of the momentum and spin of a particle. 
Moffatt~\cite{moffatt_69} adopted
this term to describe the volume integral over the scalar product of a
velocity field $\bf v$ and its curl, the vorticity $\bf \xi$ 
\begin{equation}
H = \int {\bf v} \cdot {\bf \xi}\, dV 
\end{equation}
which he named the \textit{helicity of the flow}. This idea stemmed
from the essentially kinematical interpretation of the quantity ${\bf v}
\cdot {\bf \xi}$. 
The streamlines of the flow passing near any representative point $0$ 
in a small volume element $dV$ are
(locally) helices about the streamline through $0$, and the
contribution 
\begin{equation}
{\bf v} \cdot {\bf \xi}\, dV \approx {\bf v_0} \cdot {\bf 
\xi_0}\, dV
\end{equation}
 to $H$ from $dV$ is positive or negative according as the
screw of these helices is right-handed or left-handed.    

More generally, the concept of helicity  describes the volume integral over the
scalar product of 
any general vector and its curl. In particular, the magnetic helicity
of magnetohydrodynamic flow
\begin{equation}
\label{B_helicity} 
H = \int {\bf A} \cdot {\bf B} \,dV 
\end{equation}
had already been
shown by Woltjer~\cite{woltjer_58a,woltjer_58b} to be
conserved for a perfectly conducting fluid. Later Moffatt~\cite{moffatt_69} showed that the \textit{fluid helicity} of an ideal barotropic fluid is
conserved as well.\footnote{This is true for any inviscid flow which conserves
  circulation, including irrotational flow.} Both these cases share the property that the
divergence free field lines (vortex lines or magnetic field lines)
are frozen into the fluid, and that in consequence knots 
and linkages of the field lines are inevitably conserved. It was
through struggling to understand the physical meaning of this result
that Moffatt was led to the topological interpretation of helicity in
terms of links and knots in convected vector fields generally, and to
the characterization of helicity as a \textit{topological invariant} for 
these cases. Moreau~\cite{moreau_77} observed that conservation of fluid helicity can be deduced from Noether's theorem. Helicity thus appears to have status comparable to Energy, momentum and angular momentum in this sense. As we shall 
see, unlike these quantities helicity also has a strong topological character. 
To demonstrate this we review the calculation of helicity for the simple case
of two linked flux tubes and show that the helicity counts the Gauss linking number of the tubes~\cite{moffatt_92}.

\begin{figure}[h]
\hspace{5 cm}
\vspace{1 cm}
\epsfxsize 5cm
\epsfysize 4.5cm
\epsffile{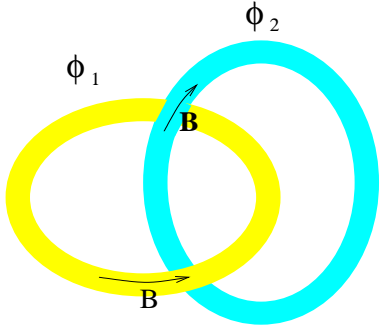}
\caption{\footnotesize{\bf Two linked $B$ flux tubes with no twist in the field
 lines and with linking number ${\cal L}_{12} = +1$.}}
\label{flux}
\end{figure}

Consider two flux tubes linked once (Fig.~\ref{flux}), carrying fluxes $\phi_1$ and $\phi_2$ respectively, of the vector field ${\bf B} = \nabla \times {\bf A} $ (the flux is zero everywhere outside the filaments).
Within each filament, the ${\bf B}$-lines are unlinked curves which close on themselves after just one passage 
round the filament. The helicity $H$ of the two flux tubes has the same form as the pseudo-scalar
quantity described in (\ref{B_helicity}).\footnote{It is straightforward that $H$
does not depend on the gauge of ${\bf  A}$; for if ${\bf A}$ is replaced
by ${\bf  A} + \nabla \psi$, then $H$ is unchanged.}

Adopting the coulomb gauge for ${\bf A}$ (i.e. $\nabla \cdot {\bf  A} = 
0$), and imposing the further condition ${\bf  A} = O(|r|^{-3})$ as $|r|
\rightarrow \infty$, ${\bf  A}$ is given by the Biot-Savart law:
\begin{equation}
{\bf  A}(r) = \frac{1}{4\pi} \int \frac{{\bf  B}(r') \times ({\bf  r} - {\bf 
r}')}{|{\bf  r}
- {\bf  r}'|^3}dV',
\end{equation}
so that from (\ref{B_helicity}),
\begin{equation}
H = \frac{1}{4\pi} \int \int \frac{{\bf  B}(r) \times {\bf  B}(r')\cdot
({\bf  r} - {\bf  r}')}{|{\bf  r}-{\bf  r}'|^3}dV dV'.
\end{equation}

 Each tube may be built of many infinitely small flux tubes carrying fluxes $\delta \phi_1$ and $\delta \phi_2$. Replacing ${\bf  B} dV$ by $\delta \phi\, d{\bf l}$: allowing for the fact that each flux filament is integrated over twice once as $r$ and once as $r'$, we
find that $\delta H = 2\, {\cal L}_{12}\,\delta \phi_1 \delta \phi_2$
with
\begin{equation}
\label{gauss_link}
{\cal L}_{12} = \frac{1}{4\pi} \oint_{C_1}\oint_{C_2}\frac{(
d{\bf l} \times d{\bf l}')\cdot ({\bf  r} - {\bf  r}')}{|{\bf  r}- {\bf  r}'|^3}
\end{equation}
which is the Gauss formula for the linking of two curves. The sign of
${\cal L}_{12}$ depends on the relative orientations of the field in 
the two filaments
(according to the right hand rule).\footnote{In this derivation, it is
essential that each flux tube should by itself have zero helicity and
this is ensured by the above assumption that the ${\bf B}$-lines within 
either tube on its own are unlinked closed curves.}

It is simple to obtain the total helicity of the two flux tubes. Each pair of filaments (one from
each tube) make a contribution $2{\cal L}_{12}\,\delta \phi_1
\delta \phi_2$ to the total helicity, so that summing
over all such pairs, this is now given by
\begin{equation}
\label{link_hell}
H = 2\,{\cal L}_{12}\,\phi_1 \phi_2.
\end{equation}

Helicity has been shown to be an invariant for flows in which the 
divergence free field lines are frozen into the flow (Woltjer~\cite{woltjer_58a,woltjer_58b} for 
a perfectly conducting fluid and Moffatt~\cite{moffatt_69} for ideal barotropic
flow). Combining this knowledge with (\ref{link_hell}), the
interpretation of the invariant in terms of conservation of 
linkage of the (vortex/magnetic) field lines which are frozen in the
fluid is immediate.

\section{\bf Helicity of Global Cosmic Strings}
\label{sec_3}
States of a Higgs field 
containing a tangle of cosmic strings resemble a tangle of quantum
vortex filaments which arise in quantum turbulent flow in superfluids.
They are both described by a complex wave function $\psi = \rho e^{\imath \theta}$ and a winding number $N$. The phase $\theta$ changes by $2\pi N$ when one 
goes once around the filament or string axes. The curve $\rho = 0$ defines
the position of the strings and the passage from $\rho = 0$ to the vacuum value
of $\rho$ takes place over a certain lengthscale denoted by $r_0$. This
lengthscale is very small in comparison with the strings lengths and distances
between the strings, i.e. the strings are very thin. 
 
Motivated by this resemblance Bekenstein~\cite{bekenstein_92} designed a helicity for cosmic strings in analogy with
fluid helicity of a superfluid.
We first review the original calculation~\cite{bekenstein_92} of the helicity
of a tangle of strings, and then correct it for the left-out contribution
of internal helicity.

\subsection{Constructing the helicity of global cosmic strings}
    
 In a superfluid the velocity is proportional
to the gradient of the phase of the wave function; Bekenstein defines the
vector 
\begin{equation}
\label{vec_v}
{\bf v} = \Theta({\rho}^2) \nabla \theta,
\end{equation}
where $\Theta({\rho}^2)$ is taken to be a function which rises rapidly from
its value at a string axis $\Theta(0) = 0$ to near its asymptotic value,
$\Theta(\rho^2_0) = 1$, over an interval of $\rho^2$ which corresponds to
a distance of order $r_0$. ${\bf v}$ is a kind of regularised velocity in that
the singularity of $\nabla \theta$ on the axis is defused by the $\Theta$
function (far from the axis ${\bf v}$ will drop inversely with the distance
from it as befits a vortex velocity field). We have
\begin{equation}
\label{curl_v}
\nabla \times {\bf v} = \Delta({\rho}^2) \nabla {\rho}^2 \times \nabla \theta,
\end{equation} 
where $\Delta({\rho}^2)$ is the derivative of $\Theta({\rho}^2)$ with respect
to ${\rho}^2$. $\Delta({\rho}^2)$ has a rather sharp peak at a distance 
$\approx r_0$ from a string axis and the integral of $\Delta$ over all 
${\rho}^2$ is unity. 

The helicity of the field ${\bf v}$ is 
defined as
\begin{equation}
\label{h}
h = \int {\bf v} \cdot \nabla \times {\bf v}\,d^3r.
\end{equation}
Unlike ordinary fluid helicity, this one vanishes.\footnote{A helicity
  which vanishes inherently cannot be used to study the dynamics of a 
  system; only results that relate to static configurations can be
  obtained. There is no physical meaning to the conservation of a
  helicity of this kind. If one is interested in drawing conclusions
  on the dynamics of a network of strings, a helicity which does not
  vanish must be constructed, and then conditions on its conservation 
  must be found.} This is clear since
the vectors ${\bf v}$ and $\nabla \times {\bf v}$ 
are perpendicular by construction. Nevertheless, string loops (and vortex rings) can link,
and knot, and their linkage is correlated with other topological properties 
of the strings to be introduced later. 

Throughout it is  
assumed that linked and knotted loops can occur over some periods with no
intersections. Situations where there are intersecting strings
are  viewed as representing  singular moments, not generic ones.
(Simulations of the development of superfluid vortices~\cite{donn_86} and global cosmic strings~\cite{shellard_90} confirm this assumption).
Bekenstein develops from the zero helicity
a quantitative relation between the
linkage and other properties of the loops.

\subsection{Evaluation of global cosmic strings helicity}
\label{global_helicity} 
In order to evaluate the helicity explicitly for a tangle of cosmic string
loops, it is convenient to rewrite (\ref{h}) as a two-point functional of 
${\bf v}$~ (This is valid if the loops are confined to a finite region~\cite{bekenstein_92}) 
\begin{equation}
\label{helic}
h = H[{\bf v}] = \int d^3r \int d^3r'\, \frac{(\nabla \times {\bf v})
  \times(\nabla \times {\bf v})' \cdot ({\bf r}-{\bf r}')}{4\pi|{\bf r}-{\bf r}'|^3}.
\end{equation}
        
We follow Bekenstein in the following discussion of the reduction of
$H[{\bf v}]$, and the conclusions that may be drawn from it. 
For field configurations of size large compared to $r_0$, and where
all the string axis are always well apart on scale $r_0$, $H$ receives
a contribution only when each of ${\bf r}$ and ${\bf r}'$ lies very close to a
string axis. This stems from the fact that the $\Delta$ function in 
(\ref{curl_v}) confines the integrals to very near the axis. 
Hence, Bekenstein assumes $H$ is equivalent to a double line integral
taken along the string axis. In addition the cross
section of each loop is very little distorted by the presence of the
other loops and nearby portions of itself and therefore is assumed to 
possess a line symmetry. Under these assumptions and a few manipulations
Bekenstein concludes that
\begin{equation}
\label{c_flux}
\int d^3r\, \nabla \times {\bf v} = 2 \pi |N| d{\bf s}
\end{equation}  
for the contribution of a little piece of string, where $d{\bf s}$ here is the vectorial line element along the piece of 
string considered. 

Inserting (\ref{c_flux}) into (\ref{helic}) we obtain a double
contribution to helicity from each pair of loops, 
because $d{\bf s}$ may progress through 
one loop (call it $\ell_i$) while $d{\bf s}'$ does so through the other
($\ell_j$) and vice-versa. 
In addition we obtain a contribution from each string loop
alone so that
\begin{equation}
\label{hel_eq}
H[{\bf v}] = \frac{1}{2}\sum_{i \neq j} 2{\cal L}_{ij}(2\pi|N_i|)(2\pi|N_j|) + \sum_i {\cal W}_i(2\pi|N_i|)^2.
\end{equation}  
where
\begin{equation}
{\cal L}_{ij} = \frac{1}{4\pi}\oint_{\ell_i}\oint_{\ell_j}\frac{d{\bf s} \times d{\bf s}' 
  \cdot ({\bf r}-{\bf r}')}{|{\bf r}-{\bf r}'|^3}
\end{equation}
is the \textit{Gauss linking number} (\ref{gauss_link}) of the two loops $\ell_i$ and $\ell_j$ and
\begin{equation}
\label{writh}
{\cal W}_i = \frac{1}{4\pi}\oint_{\ell_i}\oint_{\ell_i}\frac{d{\bf s} \times d{\bf s}' \cdot
  ({\bf r}-{\bf r}')}{|{\bf r}-{\bf r}'|^3}
\end{equation}
is the \textit{writhing} number of loop $i$ (Fuller~\cite{fuller_71}).\footnote{Equation (\ref{hel_eq}) differs from the original one in ~\cite{bekenstein_92}
by a factor of $4\pi$ which was missed in the previous calculation.} 
While ${\cal L}_{ij}$ is a topological invariant of the linking of the string 
loops and must be an integer, ${\cal W}_i$ is not a topological invariant. It 
changes with the shape of the loop and can obtain any value.
${\cal W}_i$ measures the deviation of a loop from its plane; the
writhe of a planar loop is zero.

When all loops are confined to a finite region, $H = h = 0$, and
(\ref{hel_eq}) gives
\begin{equation}
\label{const}
0 = (2\pi)^2\sum_{i \neq j} {\cal L}_{ij}|N_i||N_j| + (2\pi)^2\sum_i {\cal W}_i N_i^2.
\end{equation}  
We thus obtain a constraint connecting the linking of a collection of
loops with their geometrical shapes. Particularly simple is the case of an isolated 
loop. Then ${\cal L}_{ij} = 0$. For such a loop we must have ${\cal W}_i = 0$ too.
According to this result \textit{an isolated un-knotted loop must be a planar loop at any
instant}~\cite{bekenstein_92}. This is hard to accept since dynamically there seems to be 
no constraint on a loop accumulating a distortion continuously in an
arbitrary direction. Furthermore simulations of the
formation and evolution of cosmic strings~\cite{vachaspati_84,shellard_90} show the 
formation of loops which are not planar! Similarly, according to (\ref{const}) 
two loops linked once may never
be planar since the contribution of their writhe to the helicity must
cancel out the contribution from their linkage. This restriction is again hard to accept. 

To obtain another constraint Bekenstein introduces the relation~\cite{white_69,fuller_71}
\begin{equation}
\label{knotting}
{\cal K} = {\cal W}   + {\cal C}.
\end{equation}
${\cal K}$ measures the knotting of the curve and can jump 
by increments of
$\pm 2$, and ${\cal C}$ is the \textit{contortion} of the loop defined by 
${\cal C} = (2\pi)^{-1}\oint\tau ds$ which integrates the torsion over the
whole loop.\footnote{\label{rem}The axis of a string can be treated as a curve in
differential geometry. 
A curve $r = X(s)$ ($s$ is the arc length) has 
$T = \dot{X}(s)$ as unit tangent vector, has a unit normal $N$, and a
unit binormal $B=T \times N$. They are connected by the Frenet-Serret 
equations
\[ \frac{d}{ds} \left( \begin{array}{c} T \\ N \\ B \end{array} \right) 
 = (\kappa B - \tau T) \times  \left( \begin{array}{c} T \\ N \\ B \end{array} \right), 
\]
where $\kappa$, a positive scalar, is called the curvature of the
curve, and $\tau$, which may take on either sign, is called the
torsion. } Inserting ${\cal W}$ from (\ref{knotting}) into (\ref{const}) gives the following
constraint connecting the contortion, knotting and linking of a
collection of loops:
\begin{equation}
\sum_i {\cal C}_i N_i^2 = \sum_{i\neq j}{\cal L}_{ij}|N_iN_j| + \sum_i {\cal K}_i N_i^2. 
\end{equation}
This relation says that the sum of contortions of 
a collection of unit winding number loops is quantized in integers~\cite{bekenstein_92}.
This implies that the contortion of any single knot is quantized in
integers: the knot cannot wiggle freely about, but is frozen
in a configuration of integer contortion. This is a very strong constraint 
on the dynamics of such a configuration, and together with our previous
observations, makes us suspect 
that the calculation has missed a term.

\subsection{Internal helicity}
The first clue of the missing term was presented by Moffatt~\cite{moffatt_81},
who notes that close inspection of an expression of the form
\begin{eqnarray}
H_{\ell_i} = \int_{\ell_i} d^3r \int_{\ell_i} d^3r' \frac{(\nabla \times {\bf v})
  \times(\nabla \times {\bf v})' \cdot ({\bf r}-{\bf r}')}{4\pi|{\bf r}-{\bf r}'|^3},\nonumber
\end{eqnarray}
integrated over a single flux loop $\ell_i$ (of an arbitrary
divergence-free vector field), shows that the 
helicity in this limit has an additional
contribution 
from pairs of points ${\bf r}$, ${\bf r}'$ separated by a distance comparable to the 
cross-sectional span of the loop. In addition to the writhe ${\cal W}$ 
(\ref{writh}), which we 
introduced earlier as the sole contribution to helicity from a single loop,
Moffatt conjectures that the close points contribute an extra term namely
\begin{equation}
H_{\ell_i} = {\cal C}_{\ell_i}\phi^2
\end{equation}
where ${\cal C}_{\ell_i}$ is the contortion of loop $i$ defined
earlier, and $\phi$ is the flux of the loop. Actually
it is easy to see that Bekenstein's calculation in Sec.~\ref{global_helicity} ignores the thickness of the string, i.e. the contribution from close pairs of points ${\bf r}$ and ${\bf r}'$.
Equation~(\ref{c_flux}) reduces a volume integral to a line integral along the core 
of the string and thus treats the string as though it where infinitely thin. Moffatt speculates that the 
additional
contribution, from close pairs of points ${\bf r}$ and ${\bf r}'$, is related to the contortion so that the helicity will be a topological
invariant of the loop, i.e., $H_{\ell_i} = ({\cal W}_{\ell_i}+{\cal C}_{\ell_i})\phi^2 ={\cal K}_{\ell_i}\phi^2$ by 
(\ref{knotting}). But, thinking this over we realize that the 
contortion is also just a property of the flux tube's (string's) axis and not of its internal
structure hence we should expect a different term to appear. 

Later Moffatt~\cite{moffatt_92} explicitly calculated the helicity
of a magnetohydrodynamic system of flux tubes to find
that his earlier speculation was not exact,
and that ${\cal K}$ is not a topological invariant after
all since its value jumps
discontinuously at inflexion points (points of zero curvature). 
In this later calculation Moffatt manipulates
the magnetohydrodynamic equation of motion of a configuration of field lines 
moving with the flow. This calculation is much too 
complicated to carry over to
the case of cosmic strings since the field equations for these are
non-linear. Hence
we will follow in the footsteps of Berger and Field~\cite{berger_84} who also 
calculate
the internal helicity due to the internal structure of a flux tube, but we will 
present a slightly different calculation, and a slightly different result.    
Berger and Field's result was that $H=T\phi^2$ where $T$ is any real
number, while we obtain the same result as Moffatt~\cite{moffatt_92} with $H =
n\phi^2$ and $n$, an integer, is the linking number of the field lines
in the flux tube. 

\begin{figure}[h]
\hspace{5.8 cm}
\epsfxsize 3cm
\epsfysize 5cm
\vspace{0.2 cm}
\epsffile{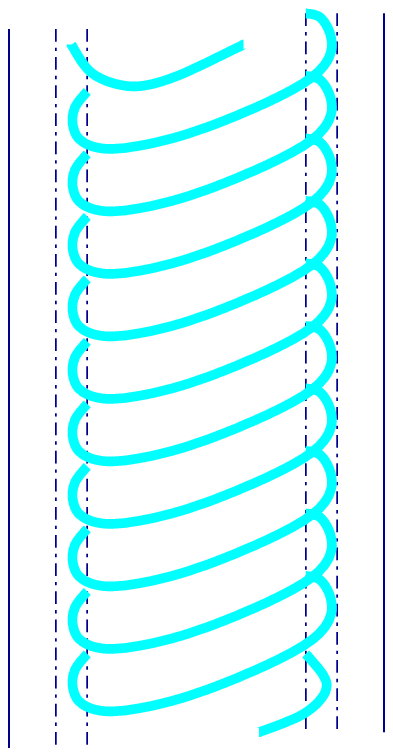}
\caption{\footnotesize{\bf The linking of flux tubes inside a string.}}
\label{f2}
\end{figure}

We first calculate exactly the contribution to the helicity of a single
vortex loop $\ell$. First we consider the inner structure of the string to consist of
nested toroidal vorticity surfaces (a peace of the loop is shown in Fig.~\ref{f2}). Then we may build up the 
string
by increments of vorticity flux $d\phi$ composed of many thin
flux tubes $\delta\phi$ that wind around the inner flux tube $\phi$ $n$
times.
The 
increment in helicity for each such flux tube $\delta\phi$ is 
$2n \phi \delta \phi$ where $n$ is the Gauss linking
number of element $\delta \phi$ with $\phi$ (the inner flux tube). The
increment in helicity for the whole of $d\phi$ is simply $2n\phi
d\phi$. (The flux $d\phi$ simply counts the number of field lines intersecting 
the increment in the cross section of the string, and hence the number
of field lines wrapped around the flux tube in the center.)
We thus have according to (\ref{helic})
\begin{eqnarray}
H_{\ell} & = & \int_{\ell} d^3r \int_{\ell} d^3r'
\frac{(\nabla\times {\bf v})\times(\nabla \times {\bf v})'
  \cdot({\bf r}-{\bf r}')}{4\pi|{\bf r}-{\bf r}'|^3} \\ \nonumber 
  & = & \int_0^{\phi}d\phi\, \phi \oint\oint \frac{(d{\bf l} \times d{\bf l}') \cdot({\bf r}-{\bf r}')}{4\pi|{\bf r}-{\bf r}'|^3} =  \int_0^{\phi}d\phi\, \phi\, 2 n
\end{eqnarray}
Where $d{\bf l}$ is along the axis of the string and $d{\bf l}'$ is along one of
the field lines in $\delta\phi$ wrapping $\phi$, and vice-versa. Finally, integrating over all the nested flux tubes
\begin{equation}
\label{int_hel}
H_{\ell} =  \frac{\phi^2}{2}2n = n(2\pi N)^2,
\end{equation}
where $N$ is the winding number of the cosmic string (the flux $\phi$ is easily
calculated using the definition of ${\bf v}$, Stoke's theorem and (\ref{q_phase})). 
 
Throughout this calculation we have assumed that $\delta\phi$ and
$\phi$ have the same linking number as a single field line in $\delta
\phi$ with the axis of the string. We have also assumed that all field
lines have the same linking number with the axis of the string (and
with each other). This is plausible for a
continuous field. As we see, the internal helicity is due to the
linking of the $\nabla \times {\bf v}$ field lines inside the string.
This linking occurs when the field $\nabla \times {\bf v}$ is twisted:
if the field lines do not link the internal helicity is zero.

We would now like to express the linking number of the field lines $n$ as a sum of ${\cal W}$ and some other
term or terms, so that we may reconcile this calculation with Bekenstein's result (\ref{hel_eq}). 
The linking number $n$ of $\delta \phi$ with $\phi$ is equal 
to the linking number
of a single field line in $\delta \phi$ with the axis of the string.
Since these are two closed curves at a very short distance apart, the
axis of the string can be treated as a curve in differential geometry,
forming the edge of a ribbon whose other edge is traced by the
field line.

This type of problem appears first to have been
addressed by C\u{a}lug\u{a}reanu~\cite{cal_59,cal_61}, who considered
two neighboring closed curves $C$, $C^*$ forming the 
boundaries of a (possibly knotted) 
ribbon of small spanwise width $\epsilon$, and showed that for sufficiently small $\epsilon$ the
linking number $n$ of $C$ and $C^*$ can be expressed 
in the form
\begin{equation}
\label{calig}
n = {\cal W} + {\cal C} + {\cal N},
\end{equation}
where ${\cal W}$ and ${\cal C}$ are respectively the writhe and
the contortion of $C$, and ${\cal N}$ is an integer representing the
number of rotations of the unit spanwise vector $U$ on the ribbon relative
to the Frenet pair (the unit normal and unit binormal\footnote{See footnote~\ref{rem}.})
in one passage round $C$. (the contortion ${\cal C}$ actually
measures the rotation round the curve of the Frenet pair itself.)
\begin{figure}[h]
\epsfxsize 8.5cm
\epsfysize 2.5cm
\hspace{3.8 cm}
\vspace{1 cm}
\epsffile{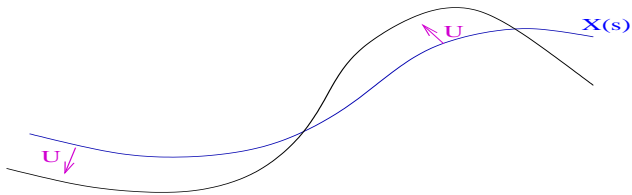}
\caption{\footnotesize{\bf Two curves form the edges of a ribbon. $U$ is the unit
  spanwise vector of the ribbon and $X(s)$ is the curve of one of the
  edges of the ribbon}}
\label{f3}
\end{figure}

The Frenet pair has discontinuous behavior in going through a point
of inflexion (zero curvature). As a result ${\cal C}$ and
${\cal N}$ are not well defined if $C$ has inflexion
points. If $C$ is continuously deformed through an inflexional
configuration, then ${\cal C}$ is discontinuous by $\pm1$, but
${\cal N}$ is simultaneously discontinuous by an equal and opposite 
amount $\mp1$, so that the sum ${\cal C} + {\cal N}$ does vary
continuously through inflexion points~\cite{moffatt_92}.
This sum is known as the \textit{twist} of the ribbon~\cite{white_69,fuller_71,fuller_78} 
\begin{equation}
\label{twist}
{\cal T} = {\cal C} + {\cal N}.
\end{equation} 
Thus, as we
remarked earlier, ${\cal K} = {\cal W} + {\cal C}$ is not a
topological invariant while $n$ (\ref{calig}) is.

With the help of (\ref{calig}) and (\ref{twist}) we are now ready to complete the 
calculation of the helicity for a single (un- knotted) cosmic string loop $\ell$. We take $C$ above to be the string's axis and $C^*$ to be along one of the field lines twisted around the axis. Then for the $i$-th loop
\begin{equation}
\label{n_hel}
H_{\ell_i} = n_i\phi^2 = {\cal W}_i(2\pi N_i)^2 + {\cal T}_i(2\pi N_i)^2.\nonumber
\end{equation}
Our final expression for the helicity of a tangle of cosmic string
loops is therefore
\begin{equation}
\label{twist_hel}
H[{\bf v}]= \sum_{i \neq j}{\cal L}_{ij}(2\pi|N_i|)(2\pi|N_j|) + \sum_i{\cal W}_i(2\pi N_i)^2 + {\cal T}_i(2\pi N_i)^2. 
\end{equation}
Contrast this (correct) result with (\ref{hel_eq}). 
The twist ${\cal T}$ is the contribution from the close pairs of points ${\bf r}$ and ${\bf r'}$ inside the string; it is not a property of the axes of the string alone, but rather of the internal structure of the field lines inside the string core.

\section{\bf New Constraints on String Configurations}
\label{sec_4}
In the previous sections we learned that helicity measures the linkage 
of the field lines of $\nabla \times \bf v$, and that the total helicity of global cosmic strings is zero whatever their configuration may be.
These properties must put some constraint on the possible configurations
of the strings. In this section we show that the knotting and
external linkage of a cosmic string determines the internal structure 
of the field lines inside the string, and thus external and internal
helicity cancel each other out so that the total helicity is zero. Since the field
lines in a string must close on themselves, and their twist depends on
external knotting and linking, we will get constraints on the possible
configurations of string loops. These constraints are weaker than those
proposed by Bekenstein, and are free from the latter's paradoxical aspects.

\subsection{Determining internal linking from winding numbers of a string loop}
Let us start by examining $\nabla \theta$ (the gradient of the phase) of 
the most simple configuration, a single loop.
As we know the phase varies when one goes once around the string by
$2\pi N$, where $N$ is
the winding number of the string.
\begin{figure}[h]
\hspace{6 cm}
\epsfxsize 4cm
\epsfysize 2.5 cm
\epsffile{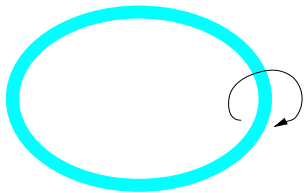}
\caption{\footnotesize{\bf Sense of variation of the phase for an isolated cosmic string loop.}}
\label{f4}
\end{figure}
 What happens if the phase varies
along
the string too? Because of periodicity this is only possible if
\begin{equation}
\oint \nabla \theta_s \cdot d{\bf s} = 2\pi N'
\end{equation}
where $d{\bf s}$ is along the string axis. But in this case there
will be another topological requirement. Because the phase varies by
$2\pi N'$ around the perimeter of the loop, for every
surface spanning the loop there must be at least one place where
the amplitude $\rho = 0$ so that the field $\psi$ will be
single-valued. This is actually the requirement that the loop be
threaded by another $N'$ unit winding-strings or a single string with
winding number $N'$. Hence an
isolated loop must have $N'=0$; for a totally
symmetric loop $\nabla \theta$ will have only one component, in the 
direction around the string.\footnote{To be more accurate, the \textbf{average} and not the value at every point, of the component of $\nabla \theta$ along the loop must be zero for an isolated loop. However, any configuration where the average of the component of $\nabla \theta$ along the loop is zero can be continuously deformed into a configuration where the component of $\nabla \theta$ along the loop equals zero at every point. Hence, as we shall see the linking of the $\nabla \times \bf v$ field lines is not affected. } 

How does this affect the configuration of 
the field lines of $\nabla \times \bf v$ ? We recall that by
definition (\ref{vec_v}), $\bf v$ is parallel to $\nabla \theta$
and $\nabla \times \bf v$ (\ref{curl_v}) is perpendicular to $\bf v$. 
Hence if $\bf v$ has a component only around the
string, $\nabla \times \bf v$ will have a component only along the
string, and its field lines will not be twisted or linked. Since linkage
of the field lines is a topological property, it will not change for
continuous deformation of the loop; therefore the field lines of any
isolated loop with arbitrary shape will be unlinked. The immediate
result is that \textit{the total helicity of any isolated loop with arbitrary 
shape is zero}. This corrects Bekenstein's conclusion
(\ref{const}) that an isolated loop must be a planar 
loop at any instant so that the helicity will be zero.

\begin{figure}[h]
\epsfxsize 8.5cm
\epsfysize 3.5cm
\hspace{2.8 cm}
\epsffile{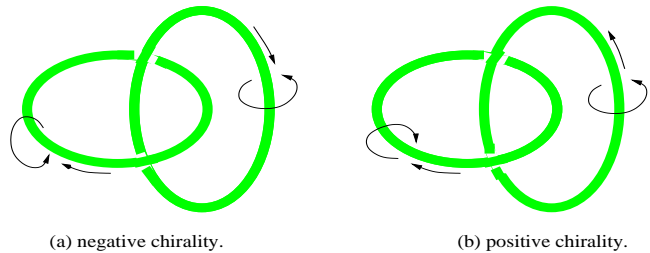}
\caption{\footnotesize{\bf Two flat linked cosmic strings. For reasons of continuity the 
  phase along the linked string must change by the same
  amount as the change in the phase around the other string. Only
  strings with the same chirality can link. }}
\label{f5}
\end{figure}

We next investigate a configuration of two linked loops $\ell_1$ and
$\ell_2$ with winding numbers $N_1$ and $N_2$ respectively. 
We now have a component of
$\nabla \theta$ in the direction along the strings too. The phase
will vary around the perimeter of loop $\ell_1$ by $2\pi N_2$ and
around the perimeter of $\ell_2$ by $2\pi N_1$ (Fig.~\ref{f5}).
 Now that $\nabla
\theta$ has components along the string and around it, we must
introduce a new concept, the \textit{chirality} (handedness) of a string loop.
A loop is right-handed (and has positive chirality) when the component of
$\nabla \theta$ along the string is in the direction of the thumb of
the right hand, whose fingers point along the component of $\nabla \theta$ around the string. If the component of
$\nabla \theta$ along the string is in an opposite direction to the 
thumb, the loop is left-handed (and the chirality is negative).

 Along with the definition of chirality comes the definition of
the sign of the directions along and around the string. We define the
positive direction around the string to coincide with the direction of
$\nabla \theta$ around it, and the positive direction along
the string to coincide with the direction of the thumb of the right
hand whose fingers follow the direction of $\nabla \theta$ around the
string axis. In a right-handed (left-handed) string loop whose phase
changes around the string by $2\pi N$ and along the string by $2\pi
N'$, $N$ and $N'$ will have equal (opposite) signs. It is easy to see that \textit{two strings may 
link only if they have the same
chirality}; otherwise there would be a discontinuity in the phase.

\begin{figure}[h]
\hspace{0.2 cm}
\hspace{4 cm}
\epsfxsize 5cm
\epsfysize 3.6cm
\epsffile{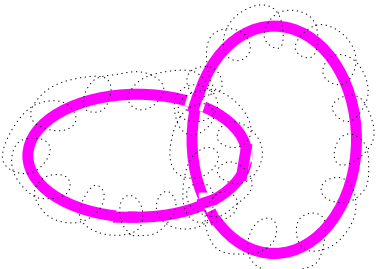}
\caption{\footnotesize{\bf The twist in the vorticity field $\nabla \times \bf v$ for
  two linked strings. When strings are linked the phase must change
  along the string too. Hence $\nabla \theta$ will have a
  component along the string in addition to its component around the
  string. As a result $\nabla \times \bf v$ also has components along and
  around the string, and its field lines are twisted and link each other.}}
\label{f6}
\end{figure}

What will be the structure of the $\nabla \times \bf v$ field lines in such a
configuration? 
Now that $\nabla \theta$ has components in two
directions, $\nabla \times \bf v$ will also have at least two components, one along the string and one around it. The field lines will be 
twisted and linked (Fig.~\ref{f6}).

In such a case we expect to have
two contributions 
to helicity, a contribution from the external linkage of the string loops,
and a contribution from the internal linkage of the field lines inside
the strings. These two contributions must cancel each other so
that this configuration may exist. The external helicity for such a
configuration with ${\cal L}_{12} = {\cal L}_{21} = \pm1$ and
winding numbers $N_1$ and $N_2$ is simply
\begin{equation}
\label{h_ext}
H_{ext} = {\cal L}_{12}(2\pi N_1)(2\pi N_2) + {\cal L}_{21}(2\pi N_2)(2\pi
N_1) = \pm2(2\pi N_1)(2\pi N_2), 
\end{equation}
where the sign of ${\cal L}_{12}$ is determined by the relative direction of
$\nabla \times \bf v$ along the perimeter of both loops. We will show below
that ${\cal L}_{12}$ is positive for right-handed
configurations and negative for left-handed ones.\footnote{Since the sign of the winding numbers ($N_1, N_2$) is always positive by our current definition, there is no need to take the absolute of the winding numbers as in (\ref{twist_hel}).}

But computing the internal helicity
\begin{equation}
\label{h_int}
H_{int} = n_1(2\pi N_1)^2 +  n_2(2\pi N_2)^2 
\end{equation}
is not straightforward. The internal
helicity of magnetic flux tubes in a magnetohydrodynamic fluid or
of vortices in an ordinary fluid is
arbitrary since there is no restriction on the linking of the field lines inside the flux tube. 
In those systems the internal structure of the field lines is not
determined by the topology of the flux tubes or vortex filaments 
and, therefore, the total
helicity cannot be determined exactly from their configuration. 
By contrast we shall now show that the internal helicity of cosmic 
strings is solely determined by the winding numbers of the strings and 
their topological configuration, so that the total helicity for a
configuration of linked vortices can be expressed as a function of their 
linkage and winding numbers alone.

\begin{figure}[h]
\hspace{4 cm}
\vspace{0.2 cm}
\epsfxsize 6cm
\epsfysize 5cm
\epsffile{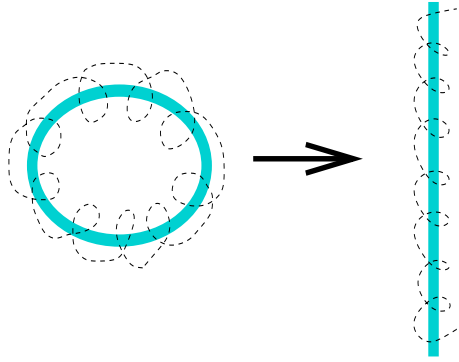}
\caption{\footnotesize{\bf A string loop may be virtually cut open and straightened out without
  changing the linking of the field lines.}}
\label{f7}
\end{figure}

We now attempt to derive the internal linking number $n$, that measures the linkage of the field lines inside
a cosmic string loop. The loop is characterized by winding numbers $N$ and $N'$ around and
along the perimeter of the loop respectively. 
Any (un-knotted) cosmic string loop may be deformed continuously into a
planar circle loop, and the twist of the field lines inside it
uniformly distributed, without changing the linking of the field lines.
This is possible since the linking is a topological
property of the field lines, and does not change by continuous deformations. 
It would further simplify our calculation if we could perform it in
cylindrical symmetric coordinates, by virtually cutting the loop and
standing it upright, such that it acquires 
cylindrical symmetry (Fig.~\ref{f7}). 
We may
convince ourselves that this is possible since very close to the axis of
the string the field lines have cylindrical symmetry around the
string, and for reasons of continuity the linkage of the field lines
cannot change as we move away from the string.   

The linking number $n$ is simply the number of times $\nabla \times
\bf v$ is wrapped around the $z$ axis (which coincides with the axis of 
the string). This must be an integer for the
string to form a loop. The vectors $\bf v$ and $\nabla \times \bf v$ 
are computed in cylindrical symmetric coordinates $(r,\varphi,z)$ as follows:
First
\begin{equation}
{\bf v} = f(r)\nabla \theta,
\end{equation}
where $\Theta({\rho}^2)$ in (\ref{vec_v}) is taken to be $f(r)$, a function of
$r$ only.
The gradient of $\theta$ must obey
\begin{equation}
\oint \nabla \theta_{\varphi} \cdot rd\varphi = 2\pi N \quad \textstyle{and} \quad \int_0^L
\nabla \theta_z \cdot dz = 2\pi N',     
\end{equation}
where $L$ is the length of the string. Hence for uniformly distributed 
twist
\begin{equation}
\nabla \theta = \frac N r \hat{\varphi} + \frac{2\pi N'} L \hat{z}.
\end{equation}
The calculation of $\nabla \times \bf v$ gives
\begin{equation}
\label{sym_curl_v}
\nabla \times {\bf v} = f^{\prime}(r)\left(N\hat{z} - \frac{2\pi r} L N'
\hat{\varphi} \right).
\end{equation}   

We are interested in the number of times each field line encircles the 
$z$ axis. Often we associate flow lines with vector fields by imagining how a test particle would move if the vector field were its velocity field. Let us assume the field line traces out the trajectory of a
point particle traveling with constant velocity during a period of
time $t$. The velocity components of the particle are equal to the components
of $\nabla \times \bf v$. The following simple equations must apply
so that the field lines may reconnect when the string forms a loop
\begin{equation}
(\nabla \times {\bf v})_z\cdot t = L \quad \textstyle{and} \quad (\nabla \times {\bf
v})_{\varphi}\cdot t = n\cdot 2\pi r.
\end{equation}
By eliminating $t$ between the two equations and using  
$\nabla \times \bf v$ from (\ref{sym_curl_v}) we obtain the
result
\begin{equation}
\label{int_link}
n = -{N'}/{N}.
\end{equation}
Thus the internal linking number $n$ depends exclusively on the winding numbers of the
loop.

\begin{figure}[h]
\hspace{1.8 cm}
\epsfxsize 11cm
\epsfysize 5cm
\vspace{0.3 cm}
\epsffile{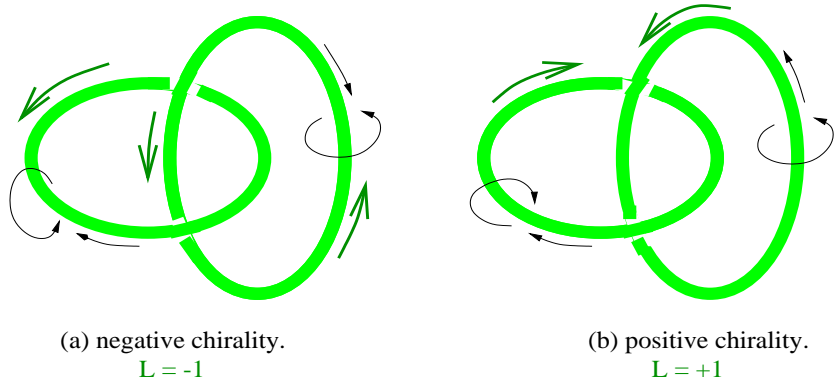}
\caption{\footnotesize{\bf Right-handed linked strings have a positive Gauss linking
  number and left-handed linked strings have a negative Gauss linking
  number. The thick arrows represent the direction of $\nabla \times
  \bf v$ along the string while the
  thin ones represent $\bf v$. $\nabla \times \bf v$ is always in
  the positive direction along the axis of the string.   }}
\label{f8}
\end{figure}

To evaluate the total helicity for two linked
un-knotted cosmic string loops, we first determine the sign of the
Gauss linking number ${\cal L}_{12}$ for left-handed and right-handed
configurations, which depends on the direction of $\nabla \times \bf
v$ along the perimeter of both loops. By our definition above, the winding
numbers around the axis of the string are always
positive, since we have defined the positive direction around the
string to coincide with the direction of $\nabla \theta$ around the
axis of the string. Hence
from (\ref{sym_curl_v}) the component of $\nabla \times \bf v$ along the 
axis is always 
in the positive direction i.e. along the thumb of the right
hand whose fingers follow the positive direction around the axis of
the string. Fig.~\ref{f8} makes 
it clear that a right-handed configuration will have ${\cal L}_{12} 
= +1$ and the left-handed configuration will have ${\cal L}_{12} 
= -1$.
The sign of the internal linking number $n$ also depends on whether
the configuration is left-handed or right-handed. As we explained
earlier, in a right-handed configuration $N$ and $N'$ have equal signs
while in a left-handed configuration, they have opposite signs.
Therefore, by (\ref{int_link}) $n$ is negative for a right-handed
pair and positive for a left-handed one. \textit{The signs of the
external and internal linkages are opposite}.

We are now able to write down the total helicity for a pair of linked strings.
For two right-handed linked strings with winding numbers $N_1$ and
$N_2$, we have $N_1^{\prime} = N_2$ and $N_2^{\prime} = N_1$ (since the axis of each loop encircles the other loop) and by
(\ref{h_ext}) and (\ref{h_int}) the total helicity is 
\begin{equation}
H = 2\cdot (2\pi N_1)\cdot (2\pi N_2) - \frac{N_2}{N_1}\cdot (2\pi {N_1})^2 - \frac{N_1}{N_2}\cdot (2\pi {N_2})^2 = 0. 
\end{equation}
For two left-handed linked loops the total helicity is
\begin{equation}
H = -2\cdot (2\pi N_1)\cdot (2\pi N_2) + \frac{N_2}{N_1}\cdot (2\pi {N_1})^2 + \frac{N_1}{N_2}\cdot (2\pi {N_2})^2 = 0. 
\end{equation}
The external and internal helicities thus cancel each other out without
constraining the shapes of the loops. 

We have thus removed the constraint on the geometrical shape of a single
loop, a pair of linked loops or any other configuration. 
The contributions to helicity are purely topological.
But (\ref{int_link}) introduces a new topological constraint. 
The internal Gauss linking number $n$ of the field lines in the string
must be an integer. Hence the winding number $N'$ along the perimeter
of the string is not free to take on just any value, i.e., a string loop with
winding number $N$ cannot be linked arbitrarily with other loops. For
the right handed pair of linked loops $\ell_1$ and $\ell_2$, for
example, the internal
linking number of $\ell_1$ is $n_1 = -N_2/N_1$ and the internal linking
number of $\ell_2$ is $n_2 = -N_1/N_2$. Therefore we must have $N_1=N_2$ 
so that both $n_1$ and $n_2$ be integers. This means that
\textit{only string loops with the same winding number may link once}.
In the next section we generalize this result to a generic configuration
of linked un-knotted loops. 

\subsection{Generalized topological constraint on configurations of
  un-knotted linked loops}
Examining a couple more examples will help us formulate a more general
result.    
\begin{figure}[h]
\hspace{4.2 cm}
\epsfxsize 6cm
\epsfysize 3.5cm
\epsffile{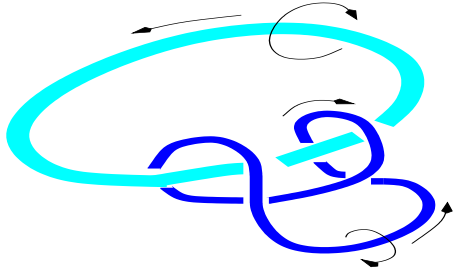}
\caption{\footnotesize{\bf Two cosmic strings linked twice.}}
\label{f9}
\end{figure}
Fig.~\ref{f9} depicts a right-handed cosmic string loop $\ell_1$
with winding number $N_1$ 
linked twice with a cosmic string loop $\ell_2$ with winding number
$N_2$ such that their Gauss linking number is ${\cal L}_{12} = 2$. 
As explained earlier the change in phase along the perimeter of a loop is 
determined by the winding numbers of the strings threading it. Since
$\ell_2$ links $\ell_1$ twice the change in phase along $\ell_1$ is
$2\pi\cdot 2N_2$ so that $N'_1 = 2N_2$; correspondingly the phase along $\ell_2$ changes by $2\pi\cdot
2N_1$ so that $N'_2 = 2N_1$.
Hence the internal linking number of $\ell_1$ is $n_1 =
-2N_2/N_1$ and that of $\ell_2$ is $n_2 = -2N_1/N_2$. Thus the values of $N_2$ must be restricted to the integers 
that divide $2N_1$ and are greater than or equal to $N1/2$. For a
few examples see Table~\ref{tab1}, only such loops may link twice.
The total helicity for this configuration is
\begin{equation}
H = 2\cdot2(2\pi N_1)(2\pi N_2) 
  - \frac{2N_2}{N_1}\cdot(2\pi N_1)^2 - \frac{2N_1}{N_2}\cdot(2\pi N_2)^2 = 0. 
\end{equation}
\begin{table}[h]
\vspace{0.5cm}
\[
\begin{array}{||l|l||} \hline
N_1   & N_2  \\ \hline
1&1,2\\
2&1,2,4\\
3&3,6\\
4&2,4,8\\
5&5,10\\
6&3,6,12\\
7&7,14\\
8&4,8,16\\
9&9,18\\
10&5,10,20\\ \hline
\end{array}
\]
\caption{\footnotesize{\bf The only permissible values of $N_2$ for different values of $N_1$
  for the configuration in Fig.~\ref{f9}.}}
\label{tab1}
\end{table}

\begin{figure}[h]
\hspace{4.5 cm}
\epsfxsize 6cm
\epsfysize 3.5cm
\epsffile{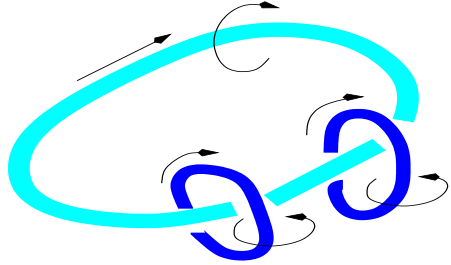}
\caption{\footnotesize{\bf A cosmic string linked once with two others.}}
\label{f10}
\end{figure}

Our second example in Fig.~\ref{f10} depicts a left-handed string loop with
winding number $N_1$ linked once with two other left-handed string
loops, one with
winding number $N_2$ and the other with winding number $N_3$.
The external Gauss linking numbers are ${\cal L}_{12} = -1$ and
${\cal L}_{13} = -1$ and ${\cal L}_{23} = 0$. The
change in phase around the perimeter of the strings, and the internal
linking numbers are determined by the same arguments as above, such
that $n_1 = (N_2 + N_3)/N_1$, $n_2 = N_1/N_2$ and $n_3 = 
N_1/N_3$. The total helicity will thus be
\begin{eqnarray}
H & =& 2\cdot (-1)(2\pi N_1)(2\pi N_2) + 2\cdot (-1)(2\pi N_1)(2\pi N_3) \nonumber \\ &+&\frac{(N_2+N_3)}{N_1}\cdot (2\pi N_1)^2 + \frac{N_1}{N_2} \cdot (2\pi N_2)^2 + \frac{N_1}{N_3}\cdot (2\pi N_3)^2 = 0
\end{eqnarray}
as we would expect.  

By generalizing from these examples we see that the internal linking
number $n_i$ for a loop $\ell_i$ linked with an arbitrary number of
other loops is 
\begin{equation}
\label{gen_int_link}
n_i = \frac{-p\cdot\sum_{j\neq i}|{\cal L}_{ij}|N_j}{N_i},
\end{equation}
where $p$ is equal to +1 for right handed loops and -1 for left handed 
ones. This equation is also a constraint on the winding numbers and
the Gauss linking numbers of the loops linking $\ell_i$, since $n_i$
must come out an integer. \newpage The total helicity of loops linked in an arbitrary 
manner thus vanishes:
\begin{eqnarray}
H & = & 4\pi^2 \left( p\cdot \sum_{i,j\neq i}
  |{\cal L}_{ij}|N_iN_j + \sum_i n_i N_i^2 \right) \\ \nonumber 
 & = & 4\pi^2 \left(p\cdot\sum_{i,j\neq i} |{\cal L}_{ij}|N_iN_j 
   - \sum_i \cdot\frac{p\cdot\sum_{j\neq
       i}|{\cal L}_{ij}|N_j}{N_i}\cdot N_i^2 \right) \\ \nonumber  
 & = & 4\pi^2 \left(p\cdot \sum_{i,j\neq i}
  |{\cal L}_{ij}|N_iN_j - p\cdot \sum_i \sum_{i\neq j}
  |{\cal L}_{ij}|N_jN_i \right ) = 0.
\end{eqnarray}

\section{\bf Electroweak Helicity of Local Cosmic Strings and Baryogenesis}
\label{sec_5}
Local cosmic strings are solutions to Lagrangians invariant under
local gauge transformations. The simplest model for a local cosmic string 
is the Abelian Higgs model; it has $U(1)$ local symmetry.
Nielsen and Olesen~\cite{nielsen_73} found straight string (cylindrically
symmetric) solutions to this model characterized by a magnetic field
confined to the string's core whose flux is quantized in proportion to
the winding number $N$ of the string. Another important property of
these strings is that they have finite energy per unit length (in contrast
to the global strings) since the gauge field $A_{\mu}$ can cancel out 
the gradient of the phase at spatial infinity.

The Nielsen-Olesen string is actually a two dimensional vortex solution 
with energy 
density localized around a point in space, which can trivially be 
embedded into the extra dimension to produce an infinite string-like object
whose energy density is localized along a line. By embedding the vortex along
a closed curve, rather than a straight line, the ends of the string can be 
joined to produce a closed string of finite length. Thus we may have a  
configuration that is a tangle of such local cosmic strings which are both knotted and 
linked with each other. Since these strings have quantized magnetic flux tubes
in their cores, we actually have a tangle of knotted and linked
magnetic flux tubes, and each configuration may be characterized by a 
magnetic helicity quantifying the linking of the cosmic strings,
and the internal linkage of the magnetic field lines inside each string.   

In this section we refer to the result~\cite{vachaspati_94,sato_95b,cohen_91,garriga_95} that the change in the electroweak magnetic helicity
of a tangle of local cosmic strings, in the standard model, is related to baryon number violation. Previous works indicated a continuous change in baryon number. In contradiction to these works, We show here, by using our knowledge of the
topological interpretation of helicity as the linking of field lines that, \textit{baryon number on local strings is quantized} in this model.

\subsection{The term $I = \int d^4x\, F_{\mu \nu}\tilde{F}^{\mu \nu}$ for Nielsen-Olesen strings}
Sato and Yahikozawa~\cite{sato_95a} scrutinize the
geometrical and topological properties of the Abelian Higgs model with 
vortex strings by calculating the topological term $I$ 
\begin{equation}
I \equiv \int d^4x\, F_{\mu \nu}\tilde{F}^{\mu \nu},
\end{equation}
which originates from the chiral anomaly, and is the CP violating $\theta$
term in the action.\footnote{See Weinberg~\cite{weinberg_95_96} ch. 22-23, Peskin \& Schroeder~\cite{peskin_95} ch. 19.} The definition of the \textit{dual} is the following: 
$\tilde{F}^{\mu \nu} = \frac{1}{2} \epsilon^{\mu\nu\rho\sigma}F^{\rho\sigma}$.
$I$ is evaluated for cases where all vortex strings are closed, with 
winding number $N=1$, and the number of the vortex strings is conserved i.e.
 vortex string reconnection does not occur. Using a 
\textit{topological formulation} which they developed, and via a 
complicated calculation, Sato and Yahikozawa show that
\begin{equation}
\label{Sato_res}
I = -\frac{8\pi^2}{e^2} \Big[ \sum_{i\neq j=1}^n {\cal L}_{ij} + \sum_{i=1}^n{\cal W}_i\Big]_{t = -\infty}^{t = +\infty}.
\end{equation}
The first term is the linking number between each pair of strings and comes 
from the \textit{mutual interactions} between different vortex strings. The
second term is the writhing number; it comes from the 
\textit{self-interactions} of each vortex string, and can vary with the shape 
of the string. 

 It is plain that the value of $I$  
changes continuously with the shape of the strings as time elapses. Equation (\ref{Sato_res}) strongly resembles
 Bekensteins result (\ref{hel_eq}) for the helicity of global cosmic strings.
We shall now show that $I$ is just the total change in time of 
the magnetic 
helicity of the Abelian
Higgs model with local vortex strings, and hence equal to the change in the
external and internal linking numbers of the magnetic field lines each weighed by the 
square of the quantized flux in the corresponding vortex.

By construction $F_{\mu \nu}\tilde{F}^{\mu \nu} = 4{\bf E} \cdot {\bf B}$,
so we may write 
\begin{equation}
\label{elec}
I = 4\int d^4\,x {\bf E} \cdot {\bf B}. 
\end{equation}
Using Maxwell's equations
\[{\bf E} = -\frac{\partial {\bf A}}{\partial t} - \nabla\phi, \quad {\bf B} = \nabla \times {\bf A} \] and \[\frac{\partial {\bf B}}{\partial t} = -\nabla \times {\bf {\bf E}}, \]
and integration by parts,
\begin{eqnarray}
\int d^4x\, {\bf E} \cdot {\bf B}& = & -\int d^3x {\bf A}\cdot {\bf B} \Big|_{t = -\infty}^{t = +\infty} + \int d^4\,x {\bf A}\cdot\frac{\partial {\bf B}}{\partial t} \\ \nonumber
 & = & -\int d^3x {\bf A}\cdot {\bf B} \Big|_{t=-\infty}^{t=+\infty} - \int d^4x\, {\bf E}\cdot \nabla \times {\bf A} 
\end{eqnarray}
where we have used the relation 
$\nabla \cdot ({\bf E} \times {\bf A}) = {\bf A}\cdot(\nabla \times {\bf E}) - {\bf E}\cdot(\nabla \times {\bf A})\,$and the assumption that all fields die off at spatial infinity.
Our final result is thus
\begin{equation}
\label{I_rel}
I = -2\int d^3\,x {\bf A}\cdot {\bf B} \Big|_{t=-\infty}^{t=+\infty} = -2\Delta H
\end{equation}
where $H$ stands for the magnetic helicity.

As we already know, the magnetic helicity may be written as the sum of the external linkages of the flux tubes and the internal linkage of the field lines
in each flux tube, times the square of the flux (\ref{twist_hel}) 
\begin{equation}
\label{mag_hel}
H = \sum_{i \neq j}{\cal L}_{ij}\cdot (\frac{2\pi}{e})^2 + \sum_i {\cal W}_i\cdot(\frac{2\pi}{e})^2 + \sum_i {\cal T}_i \cdot (\frac{2\pi}{e})^2.
\end{equation}
The two terms ${\cal W}_i + {\cal T}_i$ add to $n_i$, the internal linking number (\ref{n_hel}), and 
the magnetic flux in a local string of unit winding number in this model is $2\pi/e$.\footnote{Finite energy of the string requires $D_{\mu}\psi = (\partial_{\mu} - \imath eA_{\mu})\psi \rightarrow 0$ as $r \rightarrow \infty$. Hence, $\partial_{\mu} \theta = e A_{\mu}$ and via Stoke's theorem and (\ref{q_phase}) the magnetic flux is $2\pi / e$.}
Therefore, according to (\ref{I_rel}) and (\ref{mag_hel}) the term $I$ 
is simply equal to
\begin{equation}
\label{fin_I}
I = -\frac{8\pi^2}{e^2} \Big[\sum_{i \neq j}{\cal L}_{ij} + \sum_i {\cal W}_i + \sum_i {\cal T}_i\Big]_{t = -\infty}^{t = +\infty}.
\end{equation}
Comparing this with (\ref{Sato_res}) we see that Sato 
and Yahikozawa 
have missed out the twist term. This resulted from neglecting 
the internal 
structure of the vortex strings 
in their calculation. Our result is 
crucially different from theirs: according to their result the term $I$
may have a  continuous range of values, while according to
our result $I$ is quantized because $\Delta H$ is quantized by the internal and external linking numbers of the magnetic flux tubes threading the strings cores. This 
fact has an important physical consequence, as we shall see presently.   

\subsection{The chiral anomaly and baryogenesis}
The origin of the excess of matter over antimatter in our universe remains one
of the fundamental problems. A widely discussed model~\cite{vilenkin_94,dolgov_92,turok_95,hindmarsh_95} is  
baryogenesis in the course of the broken symmetry electroweak transition, where
the baryonic asymmetry is induced by the quantum chiral Adler-Bell-Jackiw anomaly~\cite{adler_69,bell_69}. In models of fermions coupled to gauge fields,
certain current-conservation laws are violated by this anomaly. The term $F_{\mu \nu}\tilde{F}^{\mu \nu}$ calculated on the background of the Nielsen-Olesen strings 
 has been shown to cause violation of baryon number conservation via this anomaly.

The anomaly  
arises when a classical symmetry of the Lagrangian
does not survive the process of quantization and regularization: the symmetry
of the Lagrangian is not inherited by the effective action. An axial
current that is conserved at the level of the classical equations of motion 
can thus acquire a nonzero divergence through one-loop diagrams that couple this
current to a pair of gauge boson fields. The Feynman diagram that contains this 
anomalous contribution is a triangle diagram with the axial current and the 
two gauge currents at its vertices.\footnote{A good account of this
subject can be found in Peskin~\cite{peskin_95}.} Left and right handed fermions contribute
terms with opposite signs to the anomaly; however, in chiral theories in which
the gauge bosons do not couple equally to right- and left-handed species, the 
sum of these terms can give a non-zero contribution. In theories such as QED
or QCD in which the gauge bosons couple equally to right- and left-handed 
fermions, the anomalies automatically cancel.

Within the standard model of weak interactions (the Glashow-Weinberg-Salam
$SU(2) \times U(1)$ invariant theory), the requirement from experiment that
weak interaction currents are left-handed forces us to choose a chiral gauge
coupling. The coupling of the $W$ boson to quarks and leptons can be derived
by assigning the left-handed components of quarks and leptons to doublets of
an $SU(2)$ gauge symmetry,
and then identifying the $W$ bosons as gauge fields that couple to this $SU(2)$
group, while making the right-handed fermions singlets under this group.      
The restriction of the symmetry to left-handed fields leads to the helicity
structure of the weak interactions effective Lagrangian, and thus causes the
chiral anomaly of the baryon current to arise.\footnote{All the possible gauge
anomalies of weak interaction theory must vanish for the Glashow-Wienberg-Salam
theory to be consistent. It turns out that the leptons and quarks exactly
cancel each others anomalies. In fact, the charge assignments of the quarks
and leptons in the Standard Model are precisely the ones that cancel the 
anomaly.} The baryon number 
current in 
this model is $j^{\mu}_B = \overline Q \gamma^{\mu} Q$ and the anomalous
baryonic current conservation equation is
\begin{equation}
\label{current_anomaly}                        
\partial_{\mu}j^{\mu}_B = \frac{N_F}{32 \pi^2}(-g^2
W^a_{\mu\nu}\tilde{W}^{a\mu\nu} + g'^2Y_{\mu\nu}\tilde{Y}^{\mu\nu}),
\end{equation}
where $W^a_{\mu\nu}$ and $Y_{\mu\nu}$ are the field strengths for the three 
$SU(2)$ and one $U(1)$ gauge fields $W^a_{\mu}$ ($a = 1,2,3$) and $Y_{\mu}$ respectively, and $g$ and
$g'$ are their charges. $N_F$ is the number of quark families. 

The baryon number current can be integrated to yield the change in the baryon
number between two different times, if we assume that the baryonic flux through
the surface of the three-volume of interest vanishes:
\begin{eqnarray}
Q_B &=& \int d^3x \, j^0  \\ 
\Delta Q_B &=& \int dt\,\partial_0 Q_B = \int d^4x\,\partial_0 j^0_B = \int
d^4x\,\partial_{\mu}j^{\mu}_B.
\label{delta_charge}
\end{eqnarray}
Equations (\ref{current_anomaly}) and (\ref{delta_charge}) relate the change
in the baryon number to terms of the form 
$\int d^4x\,F_{\mu \nu}\tilde{F}^{\mu \nu}$
which we calculated in the previous section for an abelian model on the 
background of Nielsen-Olesen vortices. This term was found to be related
to the change in the sum of the linking of the strings from their initial
to final configurations (\ref{fin_I}). This leads us to an exciting idea~\cite{vachaspati_94,cohen_91}: if chiral fermions
are coupled to electroweak vortices, baryon number may be violated because of the anomaly as vortices 
link and de-link. For this to happen there must be a
stable string solution in this model, and the change in the baryon number must
be related to the helicity of an electroweak field whose flux is confined
to the string.     

It is generally believed that the standard electroweak model is free from
topological defects. 
The reason is that the first homotopy group of
\[ M = G/H = \big(SU(2)_L \times U(1)\big) / U(1)_{em}\]
can be shown to be trivial i.e. $\Pi_1(M)=0$~\cite{weinberg_67,salam_68}. 
However, this does not mean
that the model is free from non-topological defects. It has been shown by
Manton~\cite{manton_83} that the configuration space of classical bosonic 
Weinberg-Salam theory has a non-contractible loop. Hence an unstable, static,
finite-energy solution of the field equations can exist.
Vachaspati~\cite{vachaspati_92,vachaspati_93} 
showed that exact vortex solutions exist, which are stable to small perturbations
for large values of the Weinberg angle $\theta_W$ and small values of the 
Higgs boson mass.\footnote{String like configurations in the Weinberg-Salam 
theory had been discussed earlier by Nambu in 1977~\cite{nambu_77}.}
 This expectation has been confirmed by
 the numerical results
of James, Perivolaropoulos and Vachaspati~\cite{james_93}.\footnote{Somewhat disappointingly, they also found that the strings are unstable for realistic parameter values
in the electroweak model. It is possible, however, that there are extensions of
the Standard Model for which the string solutions are stable.} 

Defining
\begin{eqnarray}
Z_{\mu} &=& \cos \theta_W W_{\mu}^3 - \sin \theta_W Y_{\mu} \nonumber \\
A_{\mu} &=& \sin\theta_W W_{\mu}^3 + \cos \theta_W Y_{\mu},
\end{eqnarray}
where $\tan \theta_W = g'/g$, the static vortex solution that 
extremizes the energy functional was found to be:
\begin{equation}
\label{Z_string}
\phi = f_{NO}(r)e^{im\theta} \left( \begin{array}{c} 0 \\ 1 \end{array}\right),
\quad Z_{\mu} = {A_{\mu}}_{NO},
\end{equation}
and
\begin{equation}
A_{\mu} = 0 = W_{\mu}^a (a = 1,2).
\end{equation}
The subscript $NO$ on the functions $f$ and $A_{\mu}$ means that they are
identical to the corresponding functions found by Nielsen and Olesen for
the usual Abelian-Higgs string. In other words, the Weinberg-Salam theory
has a vortex solution which is simply the $NO$ vortex of the Abelian Higgs
model embedded in the non-Abelian theory. Substituting this solution into (\ref{current_anomaly}) we have
\begin{equation}
\label{z_current}
\partial_{\mu}j^{\mu}_B = \frac{N_F \alpha^2}{32
\pi^2}\cos2\theta_WZ_{\mu\nu}\tilde{Z}^{\mu\nu}
\end{equation}
where $\alpha = g'\sin\theta_W + g\cos\theta_W$ and $\alpha^2 = g^2 + g'^2$.

The change in the baryon number is now 
\begin{equation}
\label{z_helic}
\Delta Q_B = \frac{N_F \alpha^2}{16 \pi^2}q\Delta\int d^3x
\, {\bf Z}\cdot\nabla \times  {\bf Z}
\end{equation}
via equations (\ref{delta_charge}), (\ref{z_current}) and (\ref{I_rel}) and with $q = \cos 2 \theta_W$ playing the role of baryon number charge.
We see that the change in the baryon
number is proportional to the change in the ${\bf Z}$ field's helicity. 
For a configuration of tangled strings with arbitrary winding numbers $N_i$
\begin{equation}
\Delta Q_B = \frac{N_F \alpha^2}{16 \pi^2}q \Delta \left( \sum_{i \neq
j}{\cal L}_{ij}N_iN_j + \sum_in_iN_i^2 \right)\cdot\left( \frac{4
\pi}{\alpha}\right)^2
\end{equation}
where the flux of each vortex is $4\pi N_i/\alpha$.\footnote{The
covariant derivative for this model is
\[ D_{\mu} = \partial_{\mu} + \imath\frac{\alpha}{2}Z_{\mu} \]
and hence the factor of $2$ for the flux.} 
The final result for a configuration of strings with unit winding number is
\begin{equation}
\Delta Q_B = N_Fq \Delta \left( \sum_{i \neq j}{\cal L}_{ij} + \sum_i n_i \right)
\end{equation}
and baryon number is quantized.

\section{Summary}
Earlier works failed to fully implement the topological interpretation of helicity as the measure of the linkage of field lines of the divergence free field. Since field lines may be twisted and linked inside a vortex core, the inner structure of the field inside the core may not be ignored when calculating helicity terms. Neglecting internal helicity had led to peculiar results: unacceptable constraints on string configurations and continuous violation of baryon number. Once we add the contribution from the internal structure of the strings, we find new constraints on the configurations of linked global cosmic strings, which are physically more pleasing, and we also find that, the baryon number is quantized as we would expect.

Our work is another demonstration of the fact that helicity counts the
linkage of the divergence free field lines.
We explain how there is no contradiction between the fact that helicity counts the linkage of field lines, and that the total helicity of any configuration of
knotted linked loops always manages to vanish. The vector $\mathbf v$ (\ref{vec_v}) constructed for the purpose of obtaining a helicity term for 
 global cosmic strings is proportional to $\nabla
\theta$. Hence it is perpendicular to $\nabla \times \mathbf v$ (\ref{curl_v}) and the
helicity must vanish. On the other hand the helicity counts the
linkage of field lines of $\nabla \times \mathbf v$, and these lines must
have external and internal linkages for linked configurations of cosmic 
string loops. What we have discovered is that the field lines arrange
themselves in such a manner that the external and internal helicities 
exactly cancel each other. The field lines inside a string may not link in an arbitrary manner, their linkage is constrained by the topological configuration of the string and by its linking relations to other strings. Further, we have discovered a constraint on the
permitted configurations of un-knotted linked loops, in the form of
(\ref{gen_int_link}) which must produce integer values. This constraint results from the combination of the topological character of helicity integrals with the topological nature of cosmic strings as topological defects.

In this work we confined ourselves to the problem of un-knotted linked loops.
The problem of
knotted configurations is far more complicated since
it is hard to separate the external contribution to helicity
(depending on the
topology of the knot) from the internal contribution depending on the
twist of the field lines inside the knot. Moffatt~\cite{moffatt_92}
proposed a method for calculating the external helicity, but it turns
out that for some knotted configurations this method is ambiguous (it 
may be that it works only for \textit{chiral} knots, i.e. knots that
are not isotopic to their mirror images). Hence, we leave the
treatment of knots to later work.

For local cosmic strings in the Standard Model, the change in the helicity of the electroweak magnetic field confined to the strings core is related to the change in baryon number over a period of time (\ref{z_helic}). Thus, baryon number conservation is violated
as cosmic strings link and de-link. The idea that violation of baryon number may be related to the change in the
linking of electroweak strings has been developed by 
Cohen and Manohar~\cite{cohen_91}, Vachaspati and Field~\cite{vachaspati_94} and
Garriga and Vachapati~\cite{garriga_95}. However, they neglected the contribution
of the writhe of the strings to helicity. Sato~\cite{sato_95b} adds the writhe
term, but misses out the twist term. This led to his conclusion that baryon
number conservation is violated as vortices change their shape, which implies that baryon
number may change continuously and, therefore, is not quantized. Only by
including both the writhe and the twist terms, which together give the internal
linkage of the field lines, do we recover a quantized baryon number!

Charge quantization has previously been related to
flux quantization. It was suggested by 
Dirac in 1931~\cite{dirac} 
to explain the quantization of the electric charge. This idea was developed
further by Jehle~\cite{jehle}.    
In general, quantization of charges is not a direct consequence of standard (non-GUT) field theories. Particularly, the quantization of 
electric charge has no widely accepted explanation within the 
Standard Model. During the last decade it has been suggested~\cite{foot}
that the electric charges can be heavily constrained within the framework 
of the Standard Model. This is achieved partly by constraints related to the classical structure of the theory (such as
the requirement that the Lagrangian be gauge-invariant), and from the 
cancelation of anomalies, at the quantum level. The novel idea in this work is
that quantization of the baryon number may arise from the topological nature of helicity
integrals calculated on the background of local cosmic strings and from the flux quantization in the strings.

\section*{Acknowledgments}
I thank
Prof.~Jacob.~D.~Bekenstein for his guidance during the course of this work and for many discussions. I am also grateful  
to Dr.~L.Sriramkumar for many stimulating discussions. 
This research is supported by a grant from the Israel Science Foundation,
established by the Israel Academy of Sciences and Humanities.

\end{document}